\documentclass{ifacconf}

\usepackage{graphicx}
\usepackage{natbib}
\usepackage{amsmath}
\usepackage{amsfonts}
\usepackage{xcolor}
\usepackage{graphicx}
\usepackage{subcaption}
\usepackage{tikz}
\usetikzlibrary{arrows.meta, positioning, fit}
\usepackage{bm}

\newcommand{\Gc}{{\cal G}_{\rm c}}
\newcommand{\Gi}{{\cal G}_{\rm i}}
\renewcommand{\qed}{\hfill$\square$}
\newcommand{\R}{\mathbb{R}}

\newtheorem{theorem}{Theorem}
\newtheorem{lemma}[theorem]{Lemma}
\newtheorem{proposition}[theorem]{Proposition}

\begin{document}
\begin{frontmatter}

\title{Demographic Dependence of Vaccine Adoption under Opinion Persuasion} 

\thanks[footnoteinfo]{RD is supported by the Swiss National Science Foundation under grant nr. 200021\_215336.
LM is supported by the Australian Government through the Australian Research Council’s Discovery Projects funding scheme (project DP210103700). 
PEP is supported in part by the National Science Foundation, grant
NSF-ECCS \#2238388.}

\author[First]{Alessandro Casu},    
\author[Second]{Camilla Quaresmini},
\author[Third]{Robin Delabays}, 
\author[Fourth]{Lewis Mitchell}, and
\author[Fifth]{Philip E. Par\'e\thanksref{footnoteinfo}}

\address[First]{Control Systems Group, Dept. of Electrical Engineering, Eindhoven University of Technology, The Netherlands (email: a.casu@tue.nl)}
\address[Second]{Politecnico di Milano, Milan, Italy (email: camilla.quaresmini@polimi.it)}
\address[Third]{University of Applied Sciences of Western Switzerland (HES-SO), CH-1950 Sion, Switzerland (email: robin.delabays@hevs.ch)}
\address[Fourth]{Adelaide Data Science Centre, School of Mathematical Sciences, Adelaide University, Australia (email: lewis.mitchell@adelaide.edu.au)}
\address[Fifth]{Elmore Family School of Electrical and Computer Engineering, Purdue University, West Lafayette, IN, USA (email: philpare@purdue.edu)}

\begin{abstract}                
Inspired by contagion models of social belief formation, we develop an \emph{epistemically-informed} modeling framework, SIS-Vo, in which vaccine-related information propagates on a signed opinion network. Our model allows for heterogeneous treatment effects of policy messages across subpopulations through demographic-specific responses. We derive fixed-point characterizations of the healthy (disease-free) and endemic equilibria of this model, and obtain conditions for local stability of the healthy state in terms of the contact network and opinion-dependent vaccination capacities. Using numerical simulations, we illustrate how suitably targeted policy interventions, acting through opinion dynamics, can stabilize the epidemic process by moving the system towards the healthy regime. The SIS-Vo framework thus provides a natural basis for control-theoretic analysis of vaccination policies that remain robust even when misinformation targets specific subgroups.
\end{abstract}

\begin{keyword}
Social networks and opinion dynamics;
SIS model;
misinformation; 
vaccination;
microtargeting
\end{keyword}

\end{frontmatter}

\section{Introduction}

Vaccination is one of the most effective tools for reducing the burden of infectious diseases, yet vaccine hesitancy continues to limit uptake. Recent epidemics have shown that vaccination decisions depend not only on biological and logistical factors, but also on heterogeneous beliefs and trust in institutions across demographic groups. Public health agencies increasingly deploy targeted communication campaigns, but it remains unclear how such interventions propagate through social networks and interact with epidemic dynamics.

Vaccine hesitancy and its impact on epidemic trajectories have been studied using compartmental models with capacity constraints and hesitancy-dependent uptake \citep{leung2022hesitancy,leung2023ECCopinion,bhomick2024opinion}. Related work incorporates media influence and information flows to capture how news and social media shape risk perception and behaviour \citep{mitchell2016data}. More broadly, coupled epidemic–behaviour models analyse how disease spread is intertwined with awareness, opinion formation, and protective behaviour on networks \citep{funk2010modelling,wang2015coupled}. However, behavioural responses are often modeled as homogeneous functions of prevalence or media signals, and their dependence on social structure, demographics, and targeted policy messages is typically overlooked.

In this paper, we develop an \emph{epistemically-informed} networked model that links opinion dynamics, vaccination uptake, and epidemic spread. Building on recent work on vaccine hesitancy \citep{leung2022hesitancy,leung2023ECCopinion,bhomick2024opinion} and signed opinion dynamics \citep{she2022signed}, we propose a susceptible–infected–susceptible–vaccinated model with opinions (SIS-Vo) in which vaccination confers permanent protection. Each node represents a subpopulation characterized by a demographic vector. Opinions towards vaccination evolve over a signed interaction network induced by demographic homophily and are influenced by each subpopulation's current infection level, and in turn determine an opinion-dependent carrying capacity for vaccination. A policy message enters as a control input whose perceived impact on each subpopulation depends on its demographics, inducing heterogeneous treatment effects across groups. 

We derive fixed-point characterizations of the healthy (disease-free) and endemic equilibria of the SIS-Vo model and obtain a sufficient condition for local stability of the healthy state in terms of the contact network and the vaccination carrying capacities. Numerical simulations illustrate how targeted policy messages can shift opinions, alter local carrying capacities, and thereby influence the overall impact of the disease.

The paper is organized as follows. Section \ref{sec:epistemic} presents the epistemic foundations, and Section \ref{sec:model} introduces the SIS-Vo model. Section \ref{sec:analysis} analyses the equilibria and healthy-state stability. Section \ref{sec:numerics} reports simulation results, and Section \ref{sec:conclusions} concludes with a summary and directions for future work.

\section{Epistemic foundations of demographic-weighted control}
\label{sec:epistemic}

We develop an epistemically-informed modeling framework that
\begin{enumerate}
    \item represents identity-structured opinion influence by letting social structure and homophily over demographic vectors mediate how groups affect one another’s vaccine attitudes, connecting to \cite{fricker2007epistemic}'s account of \emph{epistemic injustice} (inspired by \cite{quaresmini2025role,villa2025epistemic}) and to sociological work on homophily \citep{mcpherson2001birds}; and
    \item models epistemic dependence in the sense of \cite{hardwig1985epistemic} by allowing certain structurally central nodes in the opinion network to function as influential, and potentially misinforming, sources.
\end{enumerate}

Concretely, each subpopulation $i$ is associated with a demographic vector $\bm d_i$, and the signed opinion network has weights $\alpha_{ij} = \bm d_i^\top \bm d_j$, so that demographic similarity shapes both the sign and the strength of opinion influence.
In our model, a policy message $\bm u$ is a vector whose components correspond to different framings: demographic vectors $\bm d_i$ encode group identities and determine both who interacts with whom and how strongly each group responds to each component of the control input $\bm u$, which can be interpreted as a form of opinion persuasion, via the term $\bm u^\top \bm d_i$ in the opinion dynamics. 
This structure generates heterogeneous responses to the same message across demographic groups and network neighbourhoods, and allows us to study how different framings of $\bm u$ can be targeted to different audiences.

Following \cite{longino2020science}, we treat epistemic authority as socially organized: credibility heuristics and institutional structures are represented in the topology and weights of the opinion network and in the demographic weighting of messages. 
As in empirical work on cultural cognition and trust in science \citep{kahan2011cultural,gauchat2012politicization,sonmez2023public,sieghart2022authority,eom2025race}, the impact of communication can thus vary systematically with both the communicator’s social position and the audience’s demographic and cultural profile.

\section{Model formulation}
\label{sec:model}

\begin{figure}
    \centering

\begin{tikzpicture}[
  state/.style={draw, circle, minimum size=14mm, font=\sffamily\bfseries},
  >=Stealth,
  shorten >=2pt, shorten <=2pt
  ]

  \node[state, fill=yellow!50] (S) at (0,0) {\Large $s_i$};
  \node[state, fill=red!60] (X) at (2.6,0) {\Large $x_i$};
  \node[state, fill=green!50] (V) at (1.3,3) {\Large $v_i$};

\draw[->, thick, bend left=20] 
  (S) to node[above] {\tiny $\sum_j \beta_{ij}$} 
  (X);

\draw[->, thick, bend left=20] 
  (S) to node[left, yshift=-2mm] {\tiny $\rho\left(1-\frac{v_i}{\kappa_i}\right)$} 
  coordinate[pos=0.5] (midSV) (V);

\draw[->, thick, bend left=20] 
  (X) to node[below] {\tiny $\gamma$} 
  (S);

  \node[draw, rounded corners=6pt, inner sep=8mm,
        fit=(S) (X) (V)] 
        (ED) {};


\node[
  draw, rounded corners=4pt, thick, fill=blue!10,
  minimum width=1cm, minimum height=1cm,
  right=2.5cm of midSV, yshift=0.5cm,
  font=\sffamily\bfseries\Large
] (OD) {$o_i$};

  \draw[->, dashed, thick] (OD.west) -- ([xshift=-1mm,yshift=-3mm]midSV);

  \draw[->, dashed, thick, bend right=20] (X) to (OD);

\end{tikzpicture}
    \vspace{-5ex}
    \caption{Node-level epidemic and epistemic dynamics of the SIS-Vo model.
    The circle compartments represent the epidemic dynamics, while the square represents the opinion dynamics. 
    As illustrated by the dashed arrows, the opinions influence the vaccination adoption and the infection severity influences the opinions. }
    \label{fig:SIS-Vo}
\end{figure}
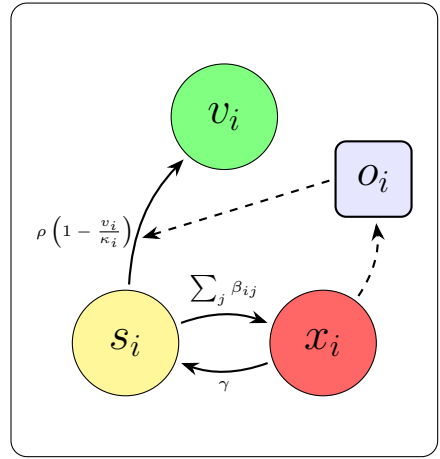

We propose a networked susceptible-infected-susceptible-vaccinated model with opinions (SIS-Vo). This model assumes that vaccinated individuals do not lose immunity. Fig.~\ref{fig:SIS-Vo} shows an overview of the dynamics of the model.

\begin{figure}
    \centering
\includegraphics[width=\linewidth]{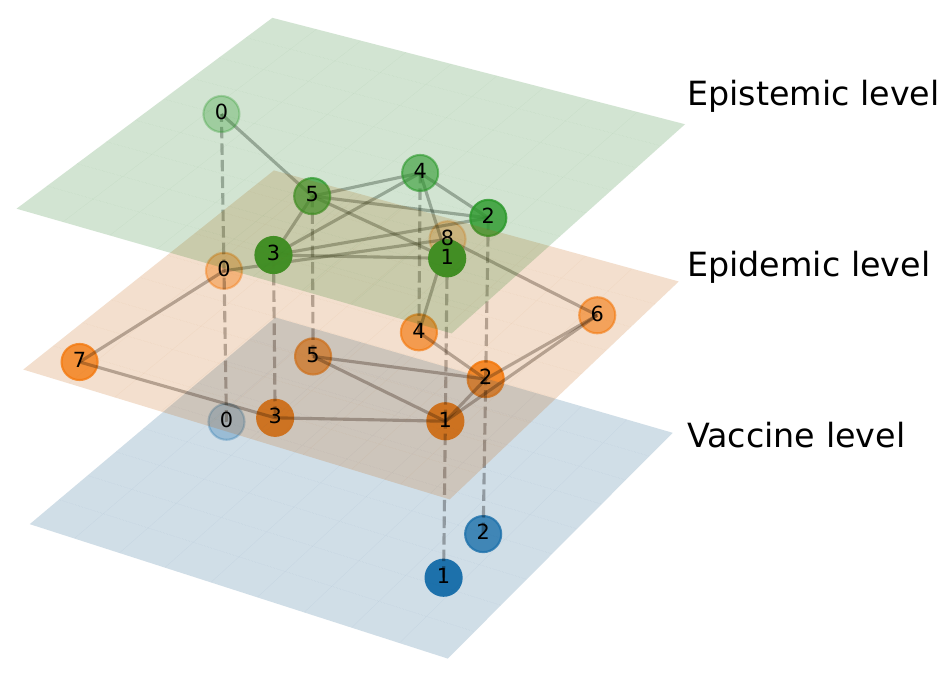}
    \caption{Multilayer network structure for the SIS-Vo model.}
    \label{fig:network_layers}
\end{figure}

Each node of the system represents a subpopulation that shares some common traits (e.g., demographics and values). 
Our SIS-Vo model describes the evolution of the epidemics and vaccination, together with the evolution of opinions, within each subpopulation. 
Fig.~\ref{fig:network_layers} shows the multi-layer network structure.

{\it Epidemic level.}
The epidemic level is modeled as a SIS-V model~\citep{leung2023ECCopinion} where each population $i$ is composed of a susceptible share $s_i\geq 0$, a contaminated share $x_i\geq 0$, and a vaccinated share $v_i\geq 0$, such that $s_i+x_i+v_i=1$. 
The infection level $x_i$ is impacted by the current infection level of subpopulation $i$ itself and by the infection levels of its neighbors in the contact network $\Gc$, whose weighted adjacency matrix is $B\in\R^{n\times n}$ is a nonnegative matrix.
Inspired by \cite{leung2022hesitancy,leung2023ECCopinion,bhomick2024opinion}, the vaccination adoption is capped by the \emph{carrying capacity} $\kappa_i$, which depends on the opinion of the subpopulation (see below). 
Altogether, the epidemic level is described by the following set of equations:
\begin{align}
    \dot{s}_i &= \gamma_i x_i - s_i \left[\sum_{j} \beta_{ij} x_j  + \rho \left(  1 - \frac{v_i}{\kappa_i}\right)  \right]\, , \label{eq:si}\\
    \dot{x}_i &= s_i\sum_{j } \beta_{ij} x_j  - \gamma_i x_i\, , \label{eq:xi}\\
    \dot{v}_i & = \rho\left( 1 - \frac{v_i}{\kappa_i}\right)s_i\, , \label{eq:vi}
\end{align}
where $\gamma_i$ is the recovery rate of subpopulation $i$, $\beta_{ij}$ is the $(i,j)$-th element of matrix $B$, and $\rho$ is the vaccination rate.

{\it Epistemic level.} 
At the epistemic level, we borrow the opinion structure from \cite{she2022signed}. 
We model the dynamics of a population's opinion towards the vaccine by a signed consensus over an interaction network $\Gi$ with weighted adjacency matrix $A\in\R^{n\times n}$ (distinct from $B$).
We include an impact of the infection level of the population $x_i$, following the rationale that vaccine adoption will increase when faced with the reality of the disease.
A population's opinion $o_i$ evolves in the interval $(-1/2,1/2)$, ranging from opposition to the vaccine ($o_i<0$) and approbation ($o_i>0$). 
The opinion dynamics is described as 
\begin{align}
    \dot{o}_i & = (x_i - o_i - \frac12) + \sum_{j} |\alpha_{ij}| (\text{sign}(\alpha_{ij}) o_j - o_i)\, , \label{eq:oi}
\end{align}
where $\alpha_{ij}$ is the $(i,j)$-th term of matrix $A$, defined as $\alpha_{ij} = \bm d_i^\top \bm d_j$.

{\it Epidemic-epistemic coupling.}
As seen in Eq.~\eqref{eq:oi}, the opinion $o_i$ is influenced by the epidemics. 
We further couple the vaccination level $v_i$ to the population's opinion $o_i$ through the carrying capacity that we choose to define as 
\begin{align}
 \kappa_i(o_i) & = \max\left(v_i,\frac{1 + 2o_i}{1 - 2o_i}\right)\, . \label{eqn:kappai}
\end{align}
The $\max$ function in Eq.~\eqref{eqn:kappai} ensures that $\dot{v}_i$ in Eq.~\eqref{eq:vi} remains nonnegative (we assume no loss of immunity after vaccination and vaccination levels should not decrease as a function of opinions) and that $\kappa_i$ remains nonnegative as well. 

Notice that our choice for dependence of $\kappa_i$ on $o_i$ is somewhat arbitrary. 
Our rationale is that we want $\kappa_i$ to be monotonous in $o_i$, taking values between $0$ (at $o_i=-1/2$) and $1$ (at $o_i=0$). 
Eq.~\eqref{eqn:kappai} is a natural choice, but not the only one.
Indeed, this choice of $\kappa_i$ implies a singularity at $o_i=1/2$. 
This singularity is not a fundamental issue for the model as it does not introduce any other singularity in other variables.

{\it Demographics.}
In order to model the discrepancies in the demographics of the subpopulations, for each subpopulation $i$, we define a \emph{demographic vector} $\bm{d}_i\in\R^k$ such that $\|\bm{d}_i\|_2=1$, which encodes $k$ distinct demographic characteristics of a population. 
Normalized vectors are considered as we do not assume any hierachies between vectors, but only different characteristics.
This demographic vector allows us to naturally define the adjacency matrix of the opinion layer, $A$, building on the notion of \emph{homophily} \citep{mcpherson2001birds}. 
Indeed, it is expected that demographically similar groups will influence each other more positively. 
From this rationale, we define the interaction between subpopulations $i$ and $j$ as $\alpha_{ij}=\bm{d}_i^\top\bm{d}_j$. 

{\it Government intervention.}
Public health agencies have the ability to broadcast advertisement messages to the population as a whole. 
Often the same message will be perceived differently by different subpopulations. 
We model this phenomenon by defining the intervention as an input unit vector $\bm{u}\in\R^k$ whose impact on subpopulation~$i$ can be interpreted as an opinion persuasion attempt and is given by $\bm{u}^\top\bm{d}_i$. 
Therefore, public agencies have the freedom to tailor their message towards one subpopulation or another, depending on their objective. 
We adapt the opinion dynamics in Eq.~\eqref{eq:oi} as
\begin{align}
    \dot{o}_i & = (x_i - o_i - \frac12) + \sum_{j} |\alpha_{ij}| (\text{sign}(\alpha_{ij}) o_j - o_i) + \bm{u}^\top\bm{d}_i \,. \label{eqn:oi-u}
\end{align}
Notice that under this modification, there is no guarantee that the opinions will remain in $(-1/2,1/2)$. 
Nevertheless, the term $\bm{u}^\top\bm{d}_i$ only introduces a shift in the opinions, which remain bounded. 

{\it Summary.}
Since $s_i+x_i+v_i=1$, one of the epidemic state equations is redundant and thus can be dropped. Therefore, the full SIS-Vo model is
\begin{align}
    \dot{x}_i &= (1-x_i-v_i)\sum_j\beta_{ij}x_j - \gamma_i x_i\, , \\
    \dot{v}_i &= \rho\left(1-\frac{v_i}{\kappa_i(o_i)}\right)(1-x_i-v_i)\, , \\
    \dot{o}_i &= (x_i-o_i-\frac12) + \sum_j\left(\alpha_{ij}o_j - |\alpha_{ij}|o_i\right) + \bm{u}^\top\bm{d}_i\, ,
\end{align}
with $\kappa_i(o_i)$ given by Eq.~\eqref{eqn:kappai}, but for simplicity we drop the dependence of $o_i$ moving forward.

\section{Model analysis}
\label{sec:analysis}
 
To identify fixed points of the model, we set $\dot{x}_i=\dot{v}_i=\dot{o}_i=0$. 
In particular, at the fixed point, $v_i^*=\min(1,\kappa_i^*)$ and $x_i^*$ satisfies
\begin{equation}\label{eq:xstar}
    x_i^* = (1-v_i^*)\frac{\sum_j \beta_{ij}x_j^*}{\gamma_i + \sum_j \beta_{ij}x_j^*}\, ,
\end{equation}
similarly to the standard networked SIS model \cite[Chap.~17]{newman2010networks}.
The carrying capacity $\kappa_i^*$ is given by Eq.~\eqref{eqn:kappai}, where the steady opinion state $o_i^*$ is given in matrix form by
\begin{equation}\label{eq:ostar}
    \bm{o}^* = (\tilde{A} + I)^{-1}\left(D^\top \bm{u} - \frac12 + \bm{x}^*\right),
\end{equation}
where
\begin{equation}
    \tilde{A}_{ij} = 
    \begin{cases}
\sum_j |\alpha_{ij}| & \text{if } i = j \\
-\alpha_{ij} & \text{if } i \ne j,
\end{cases}
\end{equation}
and $D$ consists of the vectors $\bm{d}_i$ as columns.

{\it Healthy state.}
The \emph{healthy state} where $x_i^*=0$ for all $i$ always solves Eq.~\eqref{eq:xstar}. 
In this case, the opinions $o_i^*$ (and hence, $\kappa_i^*$ and $v_i^*$) can be found in closed form from Eq.~\eqref{eq:ostar}. 
The healthy state is then of the form
\begin{align}\label{eq:healthy}
    (s_i,x_i,v_i,o_i) &= \left\{\begin{array}{ll}
        \left(0,0,1,\frac{\kappa_i^*-1}{2(\kappa_i^*+1)}\right)\, , &\text{if } \kappa_i^*\geq 1\, , \\
        \left(1-\kappa_i^*,0,\kappa_i^*,\frac{\kappa_i^*-1}{2(\kappa_i^*+1)}\right)\, , &\text{if } \kappa_i^*<1\, .
    \end{array}\right.
\end{align}

Linearizing Eq.~\eqref{eq:xi} around $\mathbf{x}^* = 0$ allows us to determine the epidemic threshold. 
The following lemma provides a sufficient condition for the stability of the healthy state.

\begin{lemma}\label{lem:epi-thresh}
    The spectrum of 
    \begin{align}
        M &= \left(I - {\rm diag}(\bm{\kappa}^*)\right)B - {\rm diag}(\bm{\gamma})
    \end{align}
is real and its largest eigenvalue is negative if, for all $i$,
    \begin{align}\label{eq:assumption1}
        \frac{\lambda_{\max}(B) - \gamma_{\min}}{\lambda_{\max}(B)} < \kappa_i^* < 1\, ,
    \end{align}
    where $\gamma_{\min}=\min_i\gamma_i$.
\end{lemma}

\begin{pf}
    For the sake of readability, let us write $C=I-{\rm diag}(\bm{\kappa}^*)$.
    The matrix 
    \begin{align}
        \tilde{M} &= C^{1/2}BC^{1/2} - {\rm diag}(\bm{\gamma})\, ,
    \end{align}
    is symmetric and similar to $M$. 
    Then by Weyl's inequality~\citep[Theorem~4.3.1]{horn2012matrix}, 
    \begin{align}
        \lambda_i(M) \leq \lambda_i(C^{1/2}BC^{1/2}) - \gamma_{\min}\, ,
    \end{align}
    and by similarity the eigenvalues of $M$ are real.

    Using the Rayleigh Quotient Theorem \cite[Theorem 4.2.2]{horn2012matrix}, for some unit vector $\bm{v}$, 
    \begin{align}
        \tilde{\lambda}_1 &= \bm{v}^\top C^{1/2}BC^{1/2}\bm{v} \leq \lambda_{\max}(B)\cdot\|C^{1/2}\bm{v}\|^2 \notag \\
        &\leq \lambda_{\max}(B)(1-\kappa_{\min}^*)\, ,
    \end{align}
    where the last inequality holds since $B$ is a nonnegative matrix and by the Perron-Frobenius Theorem~\cite[Theorem~8.4.4]{horn2012matrix}, $\lambda_{\max}(B)\geq 0$.
    Therefore, considering the eigenvalues of $M$, under the assumption \eqref{eq:assumption1},
    \begin{align}
        \lambda_1 &\leq \lambda_{\max}(B)(1-\kappa_{\min}^*) - \gamma_{\min} < 0\, ,
    \end{align}
    which concludes the proof. \qed
\end{pf}

We now have a sufficient condition for the stability of the healthy state. 

\begin{proposition}\label{prop:healthy}
    If, for all $i$, 
    $$\frac{\lambda_{\max}(B) - \gamma_{\min}}{\lambda_{\max}(B)} < \kappa_i^* < 1\, ,$$
    then the healthy state for the SIS-Vo model, defined in Eq.~\eqref{eq:healthy}, is stable.
\end{proposition}

\begin{pf}
    Linearizing Eq.~\eqref{eq:xi} in the healthy state, one sees that $\bm{x}^*=\bm{0}$ is a stable fixed point for Eq.~\eqref{eq:xi} if and only if the largest eigenvalue of $M$ is nonpositive, i.e., if Eq.~\eqref{eq:assumption1} is satisfied (Lemma~\ref{lem:epi-thresh}).
    
    In the disease-free state, $\bm{o}^*$ is computed in closed form with Eq.~\eqref{eq:ostar}. 
    This state is stable because all eigenvalues of $-\tilde{A}-I$ are nonpositive by Gershgorin's circle theorem~\cite[Theorem~6.1.1]{horn2012matrix} and any deviation in $x_i$ vanishes due to the previous discussion.

    In the disease-free state, the vaccinated share of a population $v_i$ strictly increases as long as $v_i<\kappa_i$ and $v_i<1$. 
    Indeed, $v_i$ cannot increase above $1$ and if $v_i=1$, then $\dot{v}_i=0$. 
    Furthermore, if $v_i=\kappa_i$, then $\dot{v}_i=0$. 
    In summary, in the disease-free case ($\bm{x}^*=\bm{0}$), $v_i$ increases until $\min(\kappa_i^*,1)$ and stops there. 
    Then $s_i^*$ necessarily take the value $1-v_i^*$, which concludes the proof. \qed
\end{pf}

Note that Lemma~\ref{lem:epi-thresh} and Proposition~\ref{prop:healthy} require $\bm{o} < 0$ and opinions towards vaccination are negative, as they assume $\kappa_i<1$ [see Eq.~\eqref{eqn:kappai}].
However, if $o_i > 0$, then $\kappa_i>1$ and $v_i$ increases to $1$, which is a disease-free state, indicating that the disease-free state remains stable.

{\it Endemic state.}
The set of coupled equations \eqref{eq:xstar} and \eqref{eq:ostar} can be solved numerically.
An \emph{endemic state} is found if there exists a solution with $x_i\neq 0$ for some $i$. 
The endemic equilibrium is of the form
\begin{align}
    (s_i,x_i,v_i,o_i) &= \left(1 - v_i^* - x_i^*,x_i^*,v_i^*,  \frac{\kappa_i^*-1}{2(\kappa_i^*+1)}\right)\, ,
\end{align}
where $v_i^*=\min(1,\kappa_i^*)$.
Naturally, when $\kappa_i^*\geq 1$ (i.e., $o_i^*\geq 0$), Eq.~\eqref{eq:xstar} tells us that $x_i^*=0$. 
Hence, the disease becomes endemic in subpopulations that are negatively opinionated towards vaccination, as illustrated in the simulations (see Fig.~\ref{fig:steady-state}).

{\it Intervention.}
There are various natural ways to quantify the effectiveness of an intervention $\bm{u}$.
Of course, the ideal outcome is the healthy state. 
However, as seen in Fig.~\ref{fig:intervention}, the endemic state may persist and the intervention has a strong impact on the infection levels. 

Intuitively, one may want to minimize the total number of infections in the endemic state, namely, minimizing the 1-norm of $\bm{x}^*$.
Considering that different subpopulations may be composed of  significantly different number of individuals, one may want to minimize the weighted norm
\begin{align}
    \|\bm{x}^*\|_{N,1} &= \left\|{\rm diag}(N_1,...,N_k)\cdot\bm{x}^*\right\|_1\, ,
\end{align}
where $N_i$ is the number of individuals composing subpopulation $i$. 
Notice that the vulnerabilities of different subpopulations can also easily be incorporated in the minimization by weighting each  infection level $x_i$.

Considering the nonlinear relationship between the different variables at steady state, we do not expect to find a closed form solution for infection minimization. 
Nevertheless, we can formulate the following optimization problem
\begin{align}
\begin{split}
    \underset{\bm u}{\rm minimize} & \quad \|\bm{x}^*\|_{N,1} \\
    {\rm s.t. \ \ \ } & \quad \text{Eqs.~\eqref{eqn:kappai},\, \eqref{eq:xstar},\, \eqref{eq:ostar}}\, .
\end{split}
\end{align}

An intervention is said to be aligned with the demographics of a subpopulation when it points in the same direction as the demographic vector, i.e., $\bm{u}^\top\bm{d}_i>0$. In this case, the message resonates with the cultural and social profile of that subpopulation, producing a positive drift in the opinion dynamics described by Eq.~\eqref{eqn:oi-u}. This synergy leads to larger carrying capacities $\kappa_i$, higher vaccination levels, and reduced infection prevalence in the endemic equilibrium.
Conversely, an anti-aligned intervention sets $\bm u$ in the opposite direction of the demographic vector, so that $\bm{u}^\top\bm{d}_i<0$. Such a message is poorly received in the subpopulation, shifting opinions accordingly. Fig.~\ref{fig:intervention} illustrates these effects.

\section{Simulation results}\label{sec:numerics}

\begin{figure}
    \centering
    \includegraphics[width=1\linewidth]{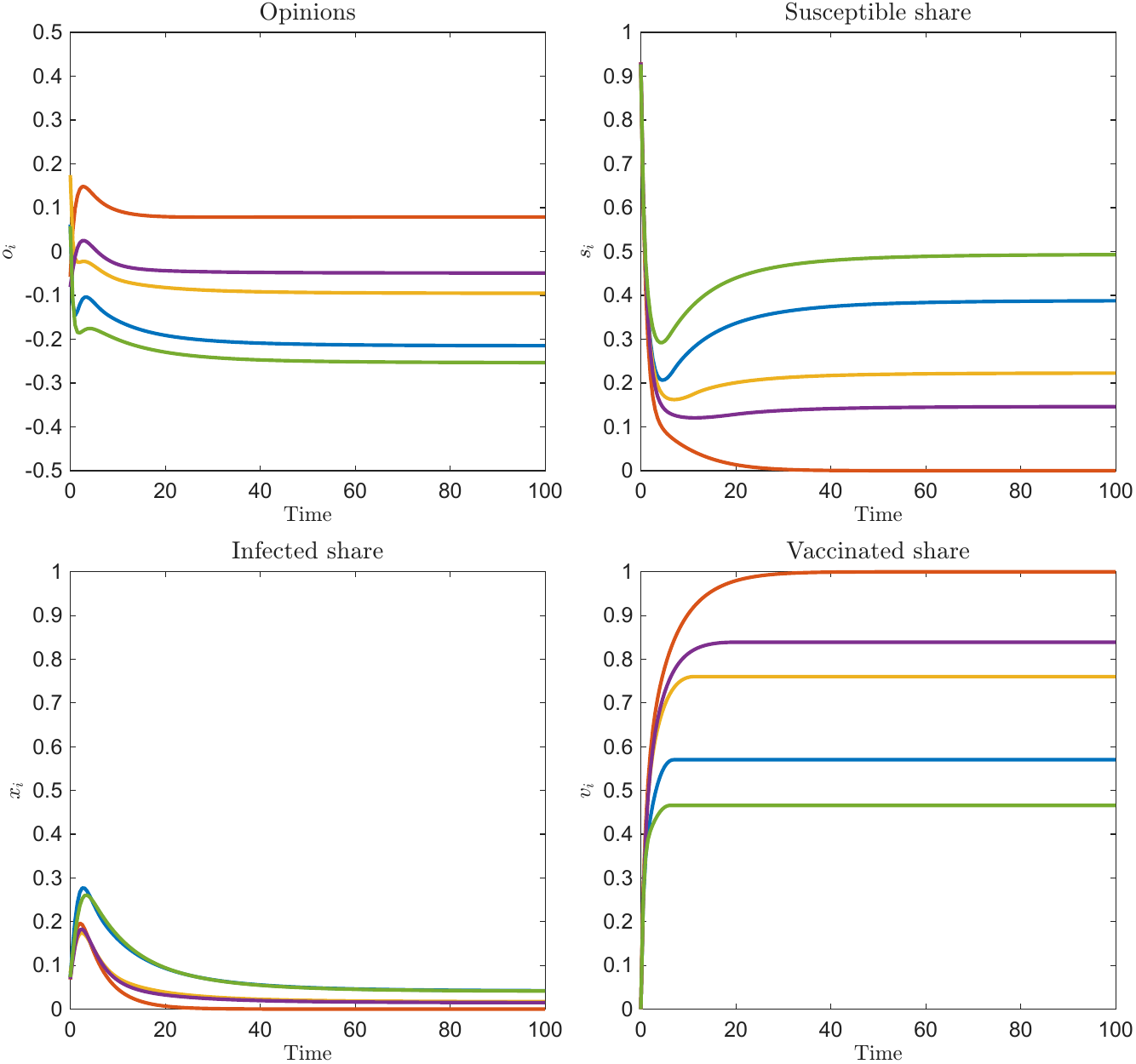}
    \caption{Example dynamics of the SIS-Vo model with 5 nodes.
    We see here an endemic state where one community (red) reaches a vanishing infection proportion and a complete vaccination, while other communities reach an endemic situation where the disease survives in the subpopulation. The choice of parameters is $\gamma=0.5$, $\rho=0.8$, a symmetric matrix with zero diagonal and off-diagonal entries between $0$ and $1$, $[\beta_{ij}]$,  randomly sampled vectors $\bm{d}_i$ (with both positive and negative entries) and an arbitrarily chosen $\bm{u}=[0.6,0.2,0.4]^\top$.}
    \label{fig:example}
\end{figure}

We illustrate the endemic behavior of the SIS-Vo model with 5 nodes in Fig.~\ref{fig:example}. Note that for the chosen parameters (including an arbitrary choice of $\bm{u}$) and randomly sampled demographic vectors $\bm{d}_i$, all but one of the nodes reach an endemic state. The only node that eradicates the virus is the one with a positive opinion state, which results in the whole subpopulation being vaccinated.

\begin{figure}
    \centering
    \includegraphics[width=\linewidth]{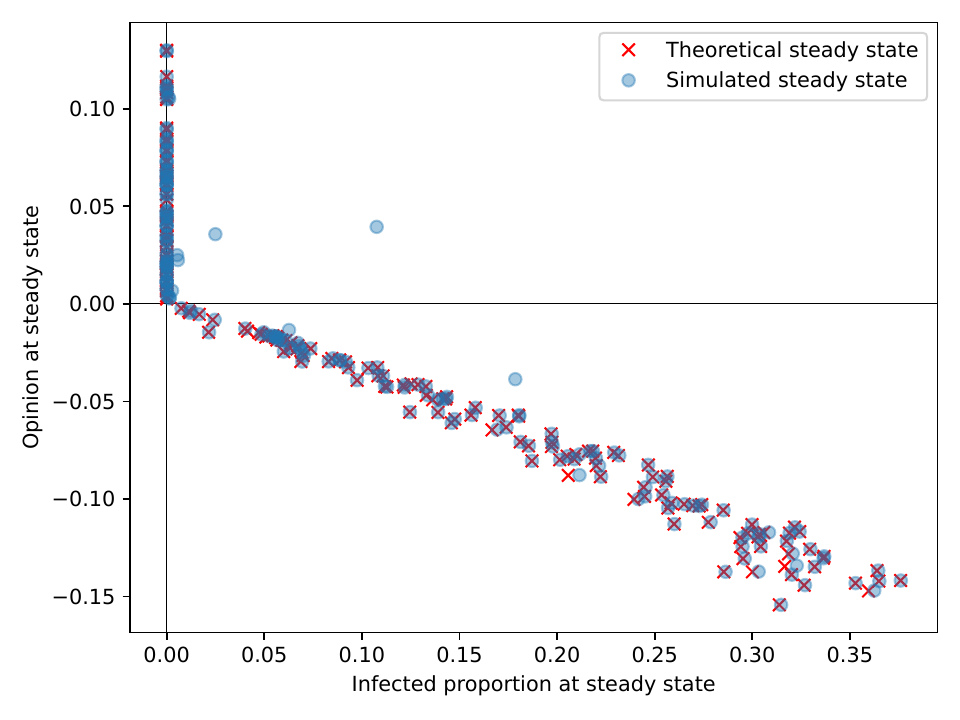}
    \caption{Steady state proportions $x_i$ versus $o_i$ for a large network ($n=200$).
    The theoretical infection proportions (red crosses) are obtained by numerically solving Eq.~\eqref{eq:xstar}. 
    In the numerical simulation, the infection proportions converge to the blue dots. 
    We observe a very good agreement between the theory and the simulations even though, due to slow convergence, some simulated infection proportions remain too high.}
    \label{fig:steady-state}
\end{figure}

To confirm that the numerical solution of \eqref{eq:xstar}-\eqref{eq:ostar} does indeed converge to the steady state solution, we simulate \eqref{eq:xi}-\eqref{eq:oi} on a large random network of 200 nodes.
We use a a single instantiation of a standard Gaussian random dot product graph with self-loops removed.
The dynamics are simulated via deterministic ODE integration of \eqref{eq:xstar}-\eqref{eq:ostar}, over 10000 timesteps with $dt = 0.2$ and $k = 4$.
Fig.~\ref{fig:steady-state} shows the simulated steady states for $x_i$ versus $o_i$, and compares with the result of 1000 steps of fixed-point iteration on \eqref{eq:xstar}-\eqref{eq:ostar}. 
Observe that agreement between simulated data and the theoretical steady states is very good, with only a small number of nodes not aligning due to slow convergence of the simulations towards the steady state.

Now using \eqref{eq:xstar}-\eqref{eq:ostar} to explore the effect of different interventions $\bm u$ on the steady state infection proportion $x_i^*$ via opinions $o_i^*$ and demographics $\bm{d}_i$, we consider three different types of interventions: 
\begin{enumerate}
    \item an intervention that is positively-aligned with demographics $u_i = {\rm sign}(\sum \bm{d}_i)\eta_i$ where $\eta_i \sim U(0,2)$,
    \item an anti-aligned intervention $u_i = -{\rm sign}(\sum \bm{d}_i)\eta_i$, and 
    \item a random intervention $u_i = \zeta_i$ where $\zeta_i \sim U(-2,2)$.
\end{enumerate}
Note that these definitions do not lead to perfectly-aligned/anti-aligned interventions; however, numerically it leads to over 80\% satisfactory cases.

\begin{figure}
    \centering
    \includegraphics[width=\linewidth]{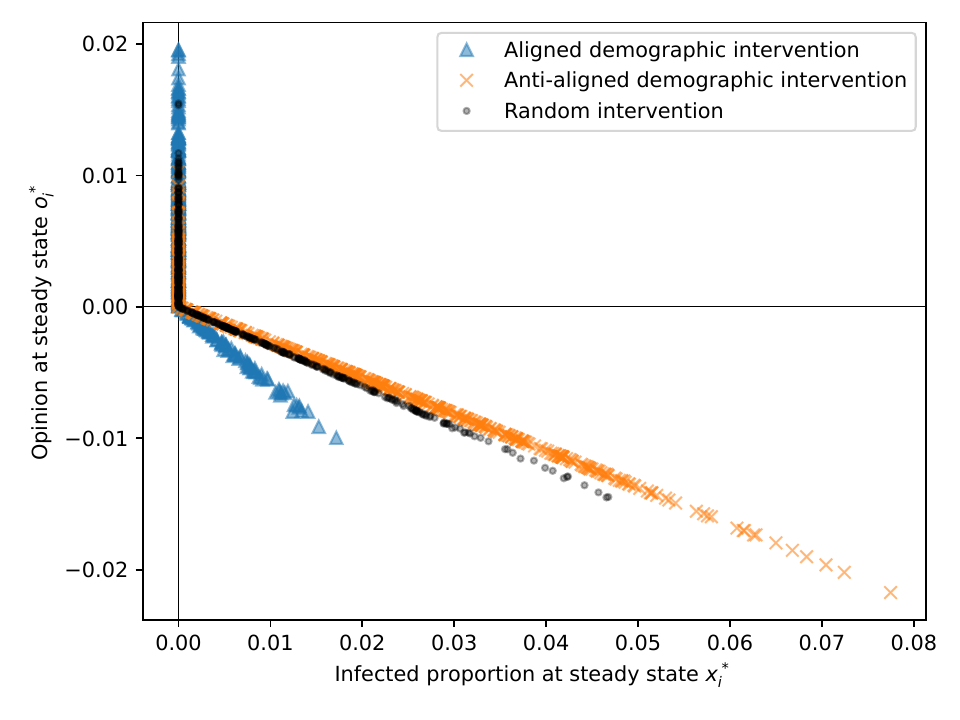}
    \caption{Opinion $o_i$ compared to the infection proportions $x_i$ under three intervention strategies $\bm{u}$. 
    The blue triangles show the outcome of when the intervention $\bm{u}$ is aligned with the demographics $\bm{d}_i$. 
    The orange crosses show the outcome when the intervention is anti-aligned with the demographics vectors. 
    The random intervention (black dots) gives an outcome that is closer to the anti-aligned strategy than to the aligned one, suggesting a strong impact of the customization of the governement's strategy.}
    \label{fig:intervention}
\end{figure}

Fig.~\ref{fig:intervention} shows the results where $n = 500$, $k = 1000$, and all other parameters as for Fig.~\ref{fig:steady-state}.
Observe that the anti-aligned intervention (orange crosses) leads to far more endemic steady states than the aligned intervention (blue triangles), and a stronger dependence of steady state infected proportion on opinions, with the random intervention (black dots) falling between the two.

Interestingly, further simulations (not shown here) confirmed that Prop.~\ref{prop:healthy} is only a sufficient condition, not necessary. 
Indeed, we were able to find some systems where the healthy state is stable, whereas Eq.~\eqref{eq:assumption1} is not satisfied. 

\section{Conclusions}
\label{sec:conclusions}
In order to model vaccine adoption in a population under general media government intervention, we propose a combination of established models of opinion dynamics and epidemic spreading. 
Opinions and disease are connected through (i) the infection level that impacts a population's opinion towards the disease, and (ii) the carrying capacity that caps the vaccination and directly depends on the population's opinion towards the disease.
Public health intervention comes in as a demographic-dependent drift in the opinion dynamics.

As preliminary results, we provide a sufficient condition for the stability of the healthy state (Prop.~\ref{prop:healthy}). 
We further explore numerically how the opinion persuasion impacts the infections in the endemic state. 
Namely, when the intervention is designed to align with the demographics' reactions, we observe a drastic reduction of the infected proportion compared to the worst-case scenario (anti-aligned intervention). 
Interestingly, random interventions appear to perform slightly better than the worst case, but still quite poorly when compared to the aligned intervention.

For future work, we envision comparing different functional relationships between the opinion $o_i$ and the carrying capacity $\kappa_i$. 
Our current choice, Eq.~\eqref{eqn:kappai} is arbitrary and it remains unclear if our results are robust against a change in this relationship.

\begin{ack}
The authors wish to acknowledge the Banff International Research Station (BIRS), where this work was conducted as part of the workshop ``Bridging the Inter-Disciplinary Gap in the Mathematical Modeling of Social Phenomena'' (25w5341). 
\end{ack}

\end{document}